\documentclass{emulateapj}
\usepackage{graphics}
\shorttitle{Cluster arcs}
\shortauthors{Ho \& White}

\begin{document}

\title{Cluster arc statistics}
\author{Shirley Ho\altaffilmark{1}, Martin White\altaffilmark{1,2}}
\altaffiltext{1}{Department of Physics, University of California, Berkeley,
CA 94720}
\altaffiltext{2}{Department of Astronomy, University of California, Berkeley,
CA 94720}

\begin{abstract}
We study the strong gravitational lensing properties of galaxy clusters
obtained from N-body simulations with standard $\Lambda$CDM cosmology.
We have used the $32$ most massive clusters from a simulation at various
redshifts and ray-traced through the clusters to investigate the giant
arcs statistics.
We have investigated the prevalence of multiple arc system, by looking
at the multiple arc fraction (defined in the paper) systematically
in various clusters and
we have found that $\sim 40-50\%$ of the clusters that produce giant arcs
give multiple arcs, which agrees with the RCS{\sc ii} observations.
We have also investigated the mass distributions that are efficient in
lensing, discussed effects of source sizes and various other factors
that are very important in the formation of giant arcs.
\end{abstract}


\section{Introduction}

Strong gravitational lensing has become an important tool in cosmology.
Lensing by clusters greatly magnifies distant sources, allowing us to
view otherwise hard to observe galaxies 
(see Metcalfe et al.~\cite{Metcalfe}; Smail et al.~\cite{Smail};
 Blain et al.~\cite{Blain}).
Giant arcs provide us a direct probe of the gravitational potential of
the lens and may enable us to study the background cosmology itself
(see Bartelmann et al.~\cite{Bart1}; Dalal et al.~\cite{Dalal};
 Macci\`o~\cite{Maccio}; Meneghetti et al.~\cite{Meneg3}).

This subject has received a great deal of attention lately because
Bartelmann et al.~\cite{Bart1} reported that the predicted number of
giant arcs varies by orders of magnitude among different cosmological
models and the observed instances of giant arcs greatly exceeded the number
of giant arcs predicted for $\Lambda$CDM (which has been widely supported
by other lines of evidence e.g., Perlmutter et al.~\cite{Perlm},
Riess et al.~\cite{Riess}, Spergel et al.~\cite{Spergel}, and
Tegmark, Zaldarriaga \& Hamilton \cite{Tegmark}).
The potential discrepancy between observations and theory in this arena is
particularly puzzling because the giant arcs are probing the matter
distribution in clusters on relatively large scales and we believe that we
understand the behavior of such dark matter dominated structures, on large
scales, very well from N-body simulations.  
This apparent discrepancy has led to significant work on refining the expected
number of giant arcs.
Bartelmann et al.~\cite{Bart2} and Meneghetti et al.~\cite{Meneg2} have
confirmed the lensing cross section predicted by
Bartelmann et al.~\cite{Bart1}.
Wambsganss et al.~\cite{Wambs} found that the lensing cross section is 
a strong function of the source redshift, making it possible to get the 
observed number of giant arcs using a broader range of source redshifts.
William et al.~\cite{Will} and Dalal et al.~\cite{Dalal} have suggested 
that massive galaxies ($10^{12}\,h^{-1}M_{\odot}$ - 
$10^{13}\,h^{-1}M_{\odot}$) are needed at the center
of most low redshift arc bearing clusters.
Further, triaxiality seems to be contributing to the formation of giant arcs,
as Oguri et al.~\cite{Oguri} found that analytical models of cluster lenses
with triaxiality and a steeper central potential enhance the lensing cross
section.  This work shows that the prediction of cross sections from theory
remains a non-trivial exercise.  In addition, in all of this work the
comparison of theory with flux and surface brightness limited observations
has been difficult.

However the activity has served to highlight the power of giant arcs to
probe cosmology and structure formation, and progress has been made in both
theory and observation.
The EMSS (Luppino et al.~\cite{Luppino}) and LCDCS
(Zaritsky \& Gonzalez \cite{Zarit}) surveys have found that the arc
frequency is $\sim 20\%$ for massive clusters.
Recent results from the Red Cluster Sequence (RCS) cluster survey suggest
a higher lensing frequency, while they also found that multiple arc systems
appear with a probability of $\sim40\%$.
New results from the ROSAT Bright Survey \cite{Schwope} 
analyzed by Kausch et al \cite{Kausch} indicates the presence of 
strong lensing events out of the 3 systems which  have  been analyzed. 
We expect that the number of giant arc systems will increase dramatically
as large optical surveys near completion.  This makes a study of giant arc
properties ever more compelling.

While it appears that some combination of the above mentioned theoretical and
observational factors may reconcile the optical depth discrepancy first
noted by Bartelmann et al.~\cite{Bart1}, other aspects of the observations
remain puzzling or controversial. 
For example, the incidence of multiple arcs or the observed redshift
distribution of lensing clusters which seems to disagree with theoretical
predictions.

Motivated by these issues we have attempted to understand how giant arcs
are made by clusters, and what they can teach us about structure formation
and cluster physics. Therefore, in this paper, we will focus on using arcs
as a way to probe clusters, and in particular why some clusters make arcs
while others don't and why multiple arcs are so prevalent
(see for example the RCS results by Gladders et al.~\cite{Gladders}).

We use numerical simulations of structure formation to generate clusters
and revisit the issues of arc formation with various different ingredients.
As previous workers have done we shall use a ray tracing technique through
dark matter halos extracted from N-body simulations.
The simulations, ray tracing methods and our results on the
dependence of cross sections on 
orientations of lensing clusters and various characteristics (sizes,
ellipticities and redshifts)  
of sources are discussed in \S\ref{sec:sims},
some of our results concerning substructure and central galaxies are
discussed in \S\ref{sec:galaxy} while our results on highly efficient
arc forming clusters are described in \S\ref{sec:multiarc}.
We conclude in \S\ref{sec:conclusions}.

\section{Simulations} \label{sec:sims}

\subsection{The N-body simulation}
                   
We wish to understand the lensing properties of clusters of galaxies which
are placed in their correct cosmological context, with a realistic
merger history and  mass
distribution and for which the intrinsic cluster properties are known.
With recent advances in N-body simulation techniques, computing power
and algorithms, this is no longer such a challenging task.
We base our work on a large, dark matter only, N-body simulation
described in (Yan, White \& Coil \cite{YanWhiCoi}; model 4).
The simulation, of a standard $\Lambda$CDM cosmology
($\Omega_{\rm m}=0.3$, $H_0=100\,h\,{\rm km}\,{\rm s}^{-1}{\rm Mpc}^{-1}$
with $h=0.7$, $\Omega_{\rm B}h^2=0.02$, $n=0.95$ and $\sigma_8=0.9$)
employs $512^3$ particles in a periodic, cubical box $256\,h^{-1}$Mpc on
a side.  This volume was chosen as a compromise between having high force
and mass resolution to resolve sub-structure in the halos and a large enough
volume to obtain several high mass clusters.
The simulation was started at $z=50$ and evolved to the present using the
{\sl TreePM\/} code described in White \cite{TreePM}.
The gravitational force softening was of a spline form, with a
``Plummer-equivalent'' softening length of $18\,h^{-1}$kpc comoving and
the particle mass is $1.04\times 10^{10}\,h^{-1}M_\odot$.

For each output we produce a halo catalog by running a ``friends-of-friends''
group finder (e.g.~Davis et al.~\cite{DEFW}) with a linking length $b=0.15$
(in units of the mean inter-particle spacing).
This procedure partitions the particles into equivalence classes, by linking
together all particle pairs separated by less than a distance $b$ which
corresponds to particles above a density of approximately
$3/(2\pi b^3)\simeq 140$ times the background density.
For each group we define the center as the minimum of the potential and
compute spherically averaged masses, velocity dispersions etc.

At each of the redshifts of interest we order the halos by mass and consider
the 32 most massive.  At $z=0$ the clusters range from
$M_{200}=1.4\times 10^{15}\,h^{-1}M_\odot$ to
$4\times 10^{14}\,h^{-1}M_\odot$
while at $z=0.7$ the range is $2-5\times 10^{14}\,h^{-1}M_\odot$.

Using the periodicity of the simulation we move each cluster to the origin
of the coordinate system and consider all of the particles within a sphere of
diameter $256\,h^{-1}$Mpc around the cluster.
This volume is large enough to contain almost all of the structure correlated
with the cluster, including filaments, merging halos etc., yet small enough
that the mass can be treated as a thin lens for sources at $z\sim 1$.
For a sequence of randomly chosen orientations we project all of the mass in
the sphere onto a $5\times 5\,h^{-1}$Mpc (comoving) grid of $512^2$ points
using a spline kernel with a smoothing equal to that of the force softening 
in the simulation.  These projected mass maps are the starting point for the
rest of the analysis.  The large number of grid points ensures that the
projected mass is smooth on the grid scale (about $10\,h^{-1}$kpc), while
the reasonably large field ensures that we are not sensitive to boundary
effects in our computations and that the tidal fields of the nearby
structures are included.

\subsection{Galaxies}

For some of the runs we include additional mass components meant to model
central (e.g.~cD) galaxies.  The galaxy is always centered on the minimum
of the cluster potential.  Specifically we add
\begin{equation}
  \Sigma = {\Sigma_0\over r(1+r^2)}
  \qquad {\rm where} \quad r=R/R_c
\end{equation}
%CHANGED
with $R=(qx^2+(y)^2/q)^{1/2}$ and $q$ the ellipticity.
We take the core radius $R_c=50\,h^{-1}$kpc.  Usually we set the mass of
the central galaxy to $1\times10^{13}\,h^{-1}M_{\odot}$.  In some cases
we add central galaxies with masses in the range 
$3\times10^{12}-3\times 10^{13}\,h^{-1}M_{\odot}$,
roughly spanning the range of observed central galaxy masses
(Sand et al.~\cite{Sand}).
Ideally we would redistribute the mass in the simulation, rather than
adding additional mass.  However the redistribution is somewhat complex
and the additional mass is so small (compared to the mass of the lensing 
cluster) that our simplification should not
matter.

\subsection{Lensing simulations}

For each projected mass map at (comoving) distance $\chi_L$ we compute the
deflection of light from a fixed source at $\chi_S$ using Fourier transform
methods making the thin lens approximation.
Within the thin lens plane a distribution of sources can be accommodated 
by using an effective value of $\chi_S$.
The rapid computation of the transforms using standard algorithms enables us
to handle the large dynamic range we desire, to ensure smooth mass
distributions on the pixel scale over large fields, while not dominating the
computing time.
The convergence is $\kappa=\Sigma/\Sigma_{\rm crit}$ where
\begin{equation}
   \Sigma_{\rm crit}= {2\over 3} L_H^2\rho_{\rm crit}
     {\chi_S\over \chi_L(\chi_S-\chi_L)} (1+z_L)^{-1}
\end{equation}
in comoving units (assuming a flat universe) and $L_H=cH_0^{-1}$ is the
Hubble length.
For a lens at $z=0.5$ and a source at $z=1$, 
$\Sigma_{\rm crit}=1.9577\times 10^{15}\,h^{-1}M_\odot/(h^{-1}{\rm Mpc})^2$.
From the $\kappa$ map we compute the deflection angle, $\alpha$, as
\begin{equation}
   \widehat{\alpha} = -i{\vec{k}\over k^2} \widehat{\kappa}
\end{equation}
where $\widehat{\phantom{\alpha}}$ indicates a Fourier transform.
We explicitly checked that this procedure works well for several analytic
potentials for which the deflections can be computed exactly.
A comparison of the caustic structure for analytic models and simulated
clusters is given in the Appendix.

To obtain the deflection angle at any point within the plane we use CIC
interpolation (Hockney \& Eastwood \cite{HocEas}) on the gridded data.
We explicitly checked that the deflections are smooth enough that this 
is a good approximation.
The computed deflection angles are typically less than an arcminute,
although on rare occasions a deflection can be as large as 1.5 arcminutes.

We then proceed in what is now the standard manner.
For a $3600^2$ grid of points in the image plane (equally spaced in angle)
we compute the source which would map to each image using the lens equation,
$\vec{\theta}_S= \vec{\theta}_I-\vec{\alpha}(\theta_I)$ where $\theta_{I}$
is the image position, $\theta_{S}$ is the source position and $\alpha$ is
the (pre-computed) deflection angle.
We scale the image and source planes to only cover the inner 1/4 of the area
of the projected mass plane.  Our tests indicated that no giant arcs formed
outside of this area.
For each source pixel we create a linked list of the image pixels to which it
maps to enable rapid calculation of images given source positions.

Our procedure is then as follows: for each map (i.e.~cluster and orientation)
we throw sources at the inner $(1/32)^2$ of the source map.  Each source is
a randomly oriented uniform ellipsoid which we produce by mapping from a
%CHANGED
circular profile with radius $s=\sqrt{qx^2+(y)^2/q}$ where $q$ is drawn
uniformly in the range $[0.5,1]$.
We shall use the term ``ellipticity'' to refer to $q$, even though with
this definition a circle has $q=1$.  We also set the semi-major axis length
to be the source size, typically $1''$, divided by $\sqrt{q}$.
The image plane is produced using the linked lists generated above and the
image is searched for giant arcs.

The arcs are found by considering all image pixels which have non-zero flux
and assigning all adjacent flux-containing pixels to the same structure.
We keep track of how many pixels are in each structure, the center of 
the structure, the pixel furthest from the center and the pixel furthest from
that pixel.  For arc-like structures we define the length as the sum of the
distances from the center to the extremal pixels described above and 
the width as the area divided by the length.
There is one issue to bear in mind with this arc finding method.
When there are two close arcs that we may visually distinguish as separate
but which contain a few connecting pixels, our algorithm will designate the
complex as one structure.
This does not happen often, however it should be kept in mind.
An ``arc'' has a length to width ratio above 7.5 and we keep track only of
those structures.

\begin{figure}
\begin{center}
\includegraphics[width=3.5in]{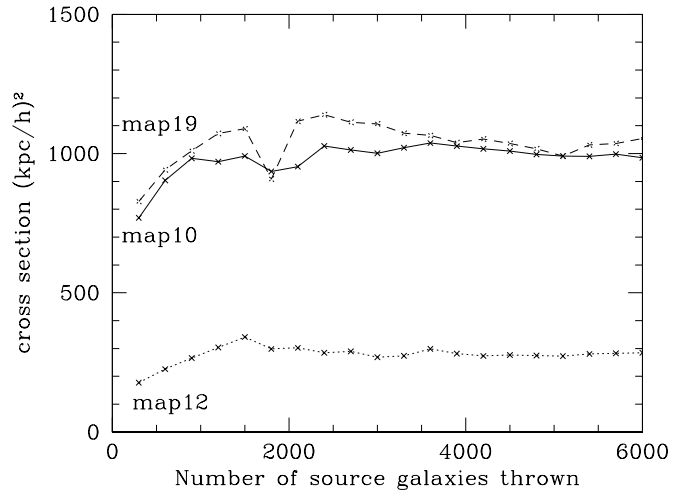}
\end{center}
\caption{The lines represent cross sections
for 3 projected maps
as a function of the number of galaxies thrown.  Convergence is
achieved after couple thousand throws.}
\label{fig:conv}
\end{figure}

We randomly throw 3 sources at a time, and repeat the procedure 800 times
per map to properly sample the source plane.
With our source size and density there is negligible chance of source overlap.
Our tests indicate that our statistical results are well converged at 800
throws (see Figure \ref{fig:conv}).
We could alternatively throw 1 source 2400 times, but the computation time
is dominated by the arc finding, so we gain efficiency by throwing
multiple sources at once.
To test convergence we threw 9000 sources for 60 different maps, and found
that the cross section achieved using 2400 sources differs from throwing 9000
sources only by $2.1\%$ on average. 
For Poisson distributed sources the cross section can be estimated from
\begin{equation}
  \sigma = A_{\rm tot}(1-e^{-\mu}){M\over N}
\end{equation}
where $A_{\rm tot}$ is the total area over which sources have been thrown, in 
our case inner $(1/32)^2$ of the source plane, $N$ is the number of sources
we have thrown, $\mu$ is the probability that a pixel is covered by a source
(i.e.~$N(A_{\rm gal}/A_{\rm tot})$, where $A_{\rm gal}$ is the area of a
galaxy) and $M$ is the number of sources which become a giant arc.

\subsection{Cross Sections}

One of the most basic quantities one can compute for any density profile
is the probability for it to form giant arcs: the cross section.
Though it is slightly off of the main focus of this work, the cross
section is easily computable from our procedure and it makes sense to
investigate it briefly.
We have found, in agreement with earlier work, that giant arcs statistics
depend on a host of factors and we shall briefly review some of them here.

First, cross section depends quite sensitively on the inner slope of the
halo density profile (Oguri et al.~\cite{Oguri}).
To illustrate this we show in Table \ref{tab:profile} the cross section as
a function of inner slope, $\beta$, for a simple analytic profile of the form
\begin{equation}
  \Sigma = {\Sigma_0\over r^{\beta}(1+r)^{3-\beta}}
    \qquad {\rm where} \quad r=R/R_c
\label{eqn:betadef}
\end{equation}
Note that the total mass here is convergent, and we have taken it to be
$1.4 \times 10^{15}\,h^{-1}M_\odot$.
We show results for $R_c=50$ and $100\,h^{-1}$kpc.
Fortunately our simulations provide guidance on the expected density profile
of the dark matter halos, and (as stated earlier) we explicitly consider the
effect of central galaxies in our calculations.  However this dependence
should be born in mind when interpreting cross-section calculations.

\begin{table}
\begin{center}
\begin{tabular}{c|cc}
$\beta$      & \multicolumn{2}{c}{cross section $(h^{-1}{\rm kpc})^2$} \\
             & $50\,h^{-1}$kpc       & $100\,h^{-1}$kpc \\
\hline
  1.00       &    36600              &   31500 \\
  1.25       &    42100              &   39700 \\
  1.35       &    44600              &   43100 \\
  1.50       &    44800              &   47100 \\
  1.75       &    47000              &   49800 \\
  1.85       &    48200              &   50200 \\
  2.00       &    48800              &   50700          
\end{tabular}
\end{center}
\caption{The dependence of the arc cross section, in $(h^{-1}{\rm kpc})^2$,
on the central slope, $\beta$, of the (analytic) density profile of
Eq.~\protect\ref{eqn:betadef}.  The first column shows results for
$R_c=50\,h^{-1}$kpc and the second for $R_c=100\,h^{-1}$kpc.}
\label{tab:profile}
\end{table}

As has been discussed in Dalal et al.~\cite{Dalal}, the giant arc 
cross section also depends strongly on the orientation of the cluster
relative to the line-of-sight.
To verify this, we have taken a few of our simulated clusters and looked at
their cross sections as a function of orientation:
a factor of 2 to 10 fluctuation in $\sigma$ is common.
Some clusters have an even stronger dependence on orientation, as we shall
discuss in \S\ref{sec:multiarc}.
The fluctuation also changed if we added central galaxies to the dark
matter maps, as such galaxies isotropize arc formation
(as will be explained later).
When we added central galaxies, the cross section fluctuations are only
changed modestly: now a factor of 2 to 8.

Second, the lensing cross section depends on the ellipticities of the sources.
For sources at $z=1.5$ and a lens at $z=0.3$ the giant arc cross sections are
$1010$, $1460$,$ 1050$ and $990$ $(h^{-1}{\rm kpc})^2$ for sources with
ellipticity $q=0.25$, $0.50$, $0.75$ and $1.0$.
This can be understood from the fact that the more elongated the sources are,
the easier it is to form elongated images.
However, there seems to be a slight preference for the dark matter runs to
form giant arcs with sources of $q=0.5$.
%This is caused by the fact that most of the caustics formed are not 
%very narrow and, as we will discuss later in the paper, most of the arcs
%form at the ends of the cluster major axis.
%Thus a source galaxy of medium ellipticity covers more of the caustics than
%a more elliptical one.

\begin{figure}
\begin{center}
\includegraphics[width=3.5in]{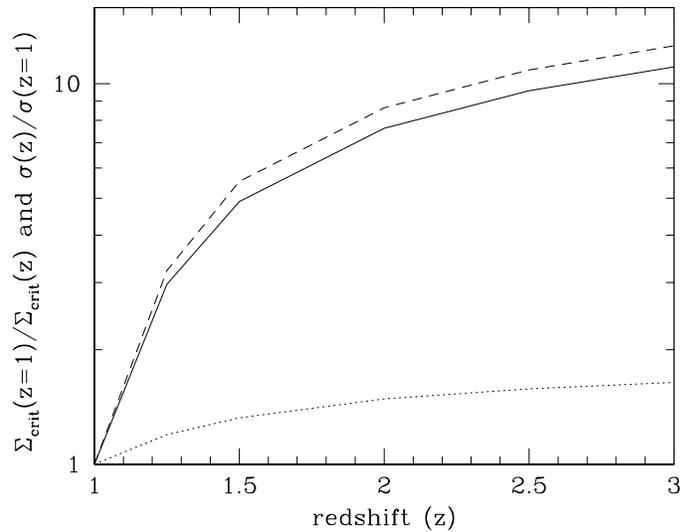}
\end{center}
\caption{The source redshift dependence of the arc statistics for a cluster
at $z=0.5$.  The lower (dotted) line shows
$\Sigma_{\rm crit}^{-1}(z)/\Sigma_{\rm crit}^{-1}(z=1)$.
The upper (dashed) line is the (normalized) cross section of giant arcs
with $L/W>10$ while the middle (solid) line is for arcs of $L/W>7.5$.}
\label{fig:red}
\end{figure}

\begin{figure}
\begin{center}
\includegraphics[width=3.5in]{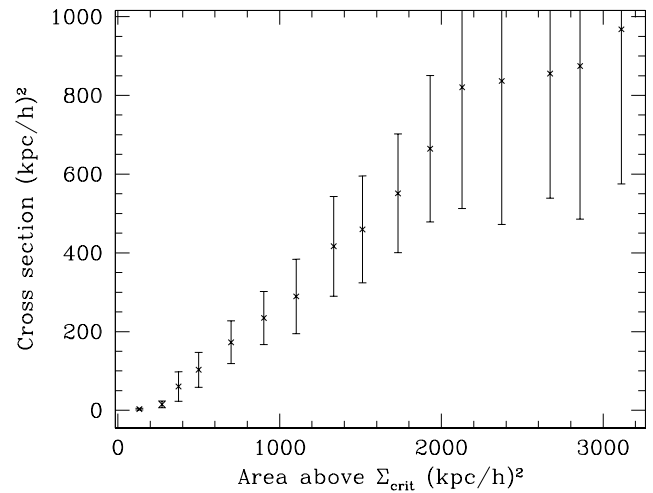}
\end{center}
\caption{The total cross section versus the total area above
$\Sigma_{\rm crit}$ for 50 maps. The mean and standard deviation of the points
are plotted in each bin.}
\label{fig:pix}
\end{figure}

Third, as it was pointed out by Wambsganss et al.~\cite{Wambs}, the lensing
cross section of the cluster is a strong function of the source redshift.
To illustrate we have taken one cluster from our simulation and put it at
$z=0.5$ while putting down the sources in the range $1<z<3$.
Figure \ref{fig:red} shows the cross section of the cluster as a function
of $z_{\rm src}$, normalized to $z_{\rm src}=1$.
For comparison we show $\Sigma_{\rm crit}^{-1}$, which determines what
fraction of the cluster mass distribution is above critical.
Note that there is an order of magnitude increase in the number of giant
arcs as we increase the source redshift, and the $z_{\rm src}$ dependence
is much stronger than simply $\Sigma_{\rm crit}^{-1}$.
We will further investigate this source redshift dependence later on from
the viewpoint of changing source sizes.
However there does seem to be a direct correlation between the area in
the cluster above $\Sigma_{\rm crit}$ and giant arc cross section, as
shown in Figure \ref{fig:pix}. Using the language of caustics, we can say that
the larger the area above $\Sigma_{\rm crit}$, the longer is the caustic,
the larger is the lensing cross section.
This is however not as simple as it seems, 
as it will be further discussed in Sec. 4.3. 

\begin{figure}
\begin{center}
\includegraphics[width=3.5in]{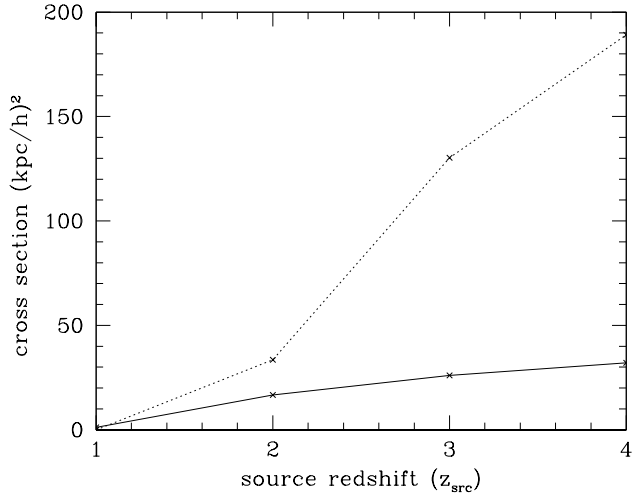}
\end{center}
\caption{The cross section versus the redshift 
averaged over 4096 projection maps. The dotted line is for  clusters
at $z=0.7$, the solid line is for clusters at
$z=0.3$.}
\label{fig:X_red}
\end{figure}

We were intrigued by the redshift distribution of lensing clusters seen
in the RCS observations (Gladders et al.~\cite{Gladders}), where essentially
all of the arc producing clusters occurred at high $z$.  Thus we investigated
how the arc cross section for clusters at different redshifts depended on
the source redshift.
For sources at relatively low redshift ($z\simeq 1$) clusters at $z=0.3$
have comparable cross sections to those at $z=0.7$.  At higher source
redshift the cross section of $z=0.7$ clusters grows more quickly than
the cross section of $z=0.3$ clusters (see Fig.~\ref{fig:X_red}).
If the majority of the sources RCS is seeing are at high redshift this
might help to explain why RCS sees arcs primarily in higher redshift
clusters.

\begin{figure}
\begin{center}
\includegraphics[width=3.5in]{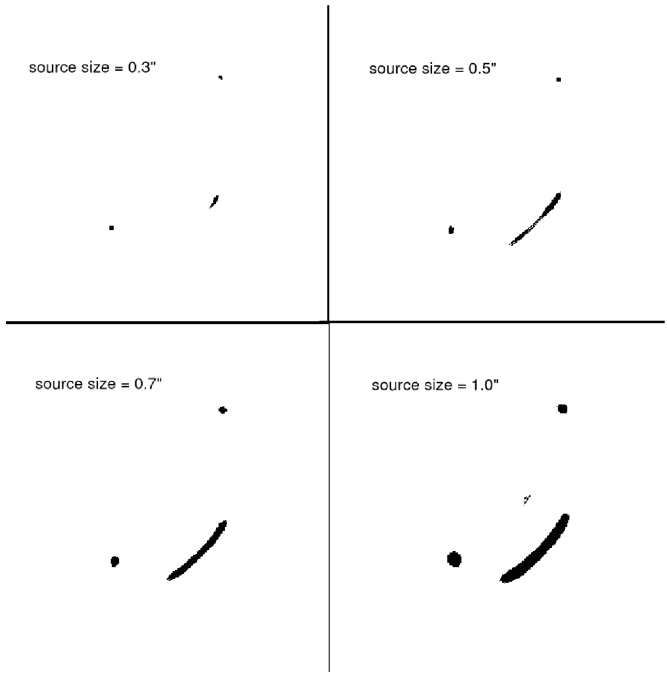}
\end{center}
\caption{Images of arcs produced by the same density map, at $z=0.3$, for
sources, at $z=2$, of four different sizes.} 
\label{fig:src}
\end{figure}

\begin{figure}
\begin{center}
\includegraphics[width=3.5in]{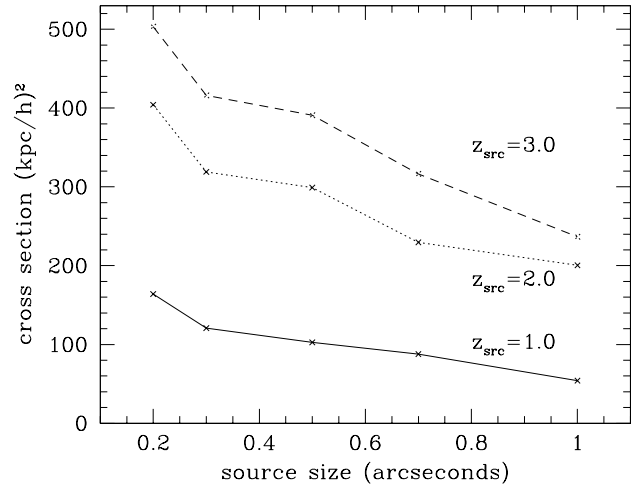}
\end{center}
\caption{The averaged cross section over 30 density maps from a cluster 
at $z=0.3$ calculated using different source sizes.}
\label{fig:Xsec_src}
\end{figure}

We also need to discuss the effect of source size\footnote{We thank the
anonymous referee for emphasizing to us the importance of this effect.}.
Contrary to the claim of Bartelmann et al.~\cite{Bart1}, we found that 
source size {\it does\/} affect the measured cross section in our simple
experiments  -- the smaller the source the larger the cross section.
As the source size is reduced, arcs get thinner (see Fig. \ref{fig:src})
and the length-to-width ratio is boosted.  Not only do the thinner arcs
look more like those in optical images, arcs with the same length but
with smaller width have larger length-to-width ratios and more arcs pass our
minimum $L/W$ cut.
This leads to a larger cross section for the same cluster when we use
smaller sources. 
To quantify this effect we have ray-traced 30 density maps with various
different source redshifts and found a consistent increase in the cross
section when we use smaller source size (see Fig. \ref{fig:Xsec_src}).
We have also tested the effect of using different source profiles,
and found that there is no significant difference between using e.g.~a
de Vaucouleurs profile and a simple constant intensity profile.

Apart from the fact that cross section changes with changing source sizes,
when we investigate the reason for the enhancement of lensing cross 
section due to the increase of source redshift 
(Wambsganss et al.~\cite{Wambs}),
we realized that common assumptions made in the lensing simulations
such as assuming a constant physical source size or constant 
angular source size will give different results that complicates 
cross comparisons between different simulations.
We found that with constant physical source size and increasing source 
redshifts, lensing
cross sections increases more dramatically than when we hold 
the angular source size constant. 
From this experiment, we can see that when we increase the source 
redshift, cross section increases not only due to the lensing weights, 
but also depends on the assumption made on the source sizes.
This can be viewed as a manifestation of the effect of source sizes too, 
as we increase source redshifts, the sources (which we assumed to have 
constant physical sizes) appear to be smaller, thus contributing to the 
increase in cross section.
We have shown the results in Table \ref{tab:redtab}
that we run with 50 maps
10 different clusters. 

\begin{table}
\begin{center}
\begin{tabular}{c|cc}
$z$      & \multicolumn{2}{c}{cross section $(h^{-1}{\rm kpc})^2$} \\
          & constant angular size & constant physical size \\
\hline
  1.00    &    0.203              &   0.203 \\
  1.25    &    1.220              &   1.627 \\
  1.50    &    4.882              &   9.153 \\
  2.00    &   15.662              &  33.358 \\
  3.00    &   33.968              &  77.597  \\
  4.00    &   46.579              &  86.445 \\
  5.00    &   54.308              & 121.633
\end{tabular}
\end{center}
\caption{Using 50 density maps taken from 10 different clusters, 
with a range of masses, varying source redshift,
along with constant angular source size 
($1''$) or
physical source size ($11.2056 h^{-1}$kpc which also translates to $1''$ at $z=1$).}
\label{tab:redtab}
\end{table}

While the cross section is somewhat sensitive to source size, we have
found that other statistics, including those related to multiple arcs
which are the main focus of this paper, are not very sensitive to the
source size.  For this reason we shall stick with a constant source
size (of $1''$) and uniform intensity disks in our modeling.
Calculations aimed at predicting the cross section should include a
realistic source size distribution and redshift distribution.

Torri et al.~\cite{Torri} suggested that the strong lensing cross section
has an interesting dependence on recent merger events.  In particular,
they have shown that merging clusters can have their cross sections
enhanced by an order of magnitude during the merging process.
This in principle could solve the discrepancy between the theoretical
prediction and the observed number of giant arcs.
However, they have used a cluster at $z\sim 0.3$ and artificially scaled
its mass from $7\times 10^{14}\,h^{-1}M_{\odot}$ to $10^{15}\,h^{-1}M_{\odot}$
with the merging structure being 25\% of the mass of the main cluster.
It is not clear if such events are common enough to explain the RCS results.
Further our studies suggest that truly efficient lenses arise when a
large, contiguous region is above the critical density.  Naively we
imagine that mergers would lead to less ``relaxed'' high density material
near the center of the cluster.
This issue deserves further investigation, but we do not have the necessary
simulations in hand at present.

As we have seen from the above discussion, the cross section of a cluster
depends on many different, difficult to model, factors.  This makes it
challenging to predict the cross section and use these predictions as a
probe of the background cosmology.  Conversely it allows us to study many
diverse phenomenon using a sample of giant arcs.

\section{Central galaxies: ellipticities and orientations} 
\label{sec:galaxy}

One frequently finds massive galaxies at the center of clusters and this
galaxy has a mass distribution which has been concentrated by the effects
of baryonic cooling.
This can affect the total mass profile in the region which is important for
giant arc formation.
Meneghetti et al.~\cite{Meneg2} have argued that central galaxies have a
substantial effect on cluster arc formation, while Dalal et al.~\cite{Dalal}
have argued that they only affect arcs at small radii -- though they
isotropize the angular distribution of the arcs.

\begin{figure}
\begin{center}
\includegraphics[width=3.5in]{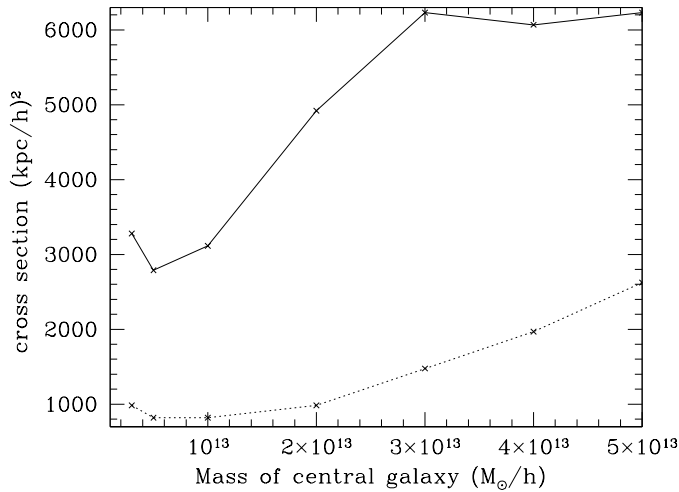}
\end{center}
\caption{The arc cross section as a function of the mass of the central
galaxy added to 50 density maps, produced from 
projected densities of 50 different orientations of a cluster with mass
$1.249\times 10^15$ $h^{-1}M_{\odot}$.
Note that when no galaxy is added, the result is
indistinguishable from that of a galaxy of $5\times10^{12}\,h^{-1}M_{\odot}$.
The solid line is the cross section for arcs with $L/W >=7.5$, the dotted
line is for arcs with $L/W >=10$.}
\label{fig:cdadd}
\end{figure}

To investigate this, we artificially added central ``galaxies'' of different
masses to the center of one of our simulated clusters.
The galaxy was modeled as a randomly oriented ellipsoid with $q$ drawn
uniformly in the range $[0.5,1]$.
Figure \ref{fig:cdadd} shows the increase in the giant arc cross section
as a function of the mass of the added galaxy.
At the higher mass end the giant arcs cross section is tripled.
However we find, in agreement with Dalal et al.~\cite{Dalal}, that the
addition of central galaxies forms arcs primarily at small radii and there
is an isotropization of the arc positions.

\begin{figure}
\begin{center}
\resizebox{3.5in}{!}{\includegraphics{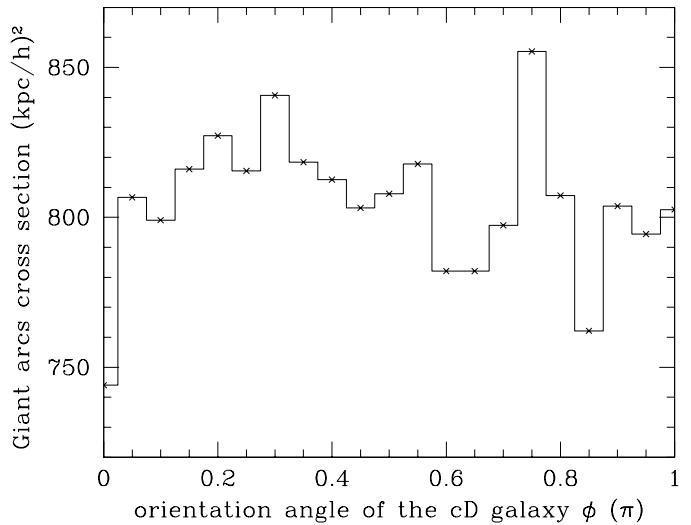}}
\end{center}
\caption{The dependence of the cross section on the orientation of the
central galaxy.  For each orientation there are 20 different ellipticities.
We plot the average of 20 runs of galaxies lying in the same direction 
but with different ellipticities.}
\label{fig:phi}
\end{figure}

\begin{figure}
\begin{center}
\resizebox{3.5in}{!}{\includegraphics{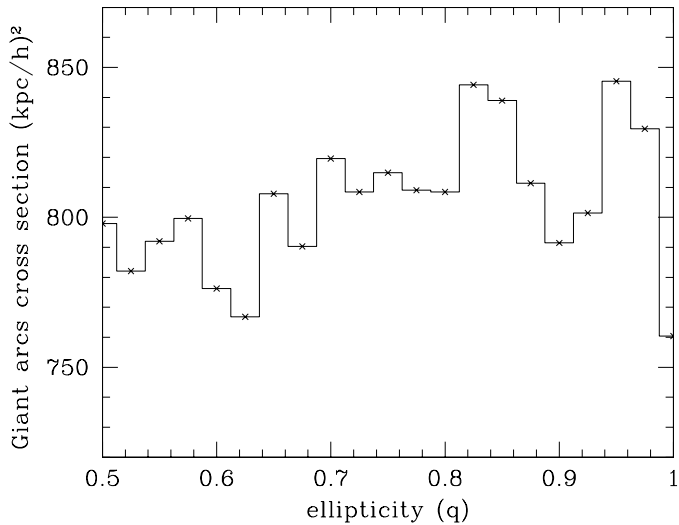}}
\end{center}
\caption{The ellipticity of the central galaxies causes the cross section of 
the cluster to fluctuate.  For each ellipticity there are 20 runs of different
orientations of the galaxy.  We plot the average of 20 galaxies with same
ellipticities but different orientations.}   
\label{fig:ellip}
\end{figure}

Figure \ref{fig:phi} shows that the lensing efficiency is much greater
for some orientations of the central galaxy than others.
In particular the efficiency is maximized when the central galaxy aligns
with the projected mass of the underlying cluster, at $\phi\simeq 0.75$
in our example.
Though the ellipticity of the central galaxy causes the cross section to
fluctuate (see Fig. \ref{fig:ellip}), 
the cross section does not seem to have a secular dependence
on the (2D) ellipticities of the galaxies. This may be because several
different factors are contributing to the effect.
Depending on the ellipticities and the ratio of the size of the host cluster
to the size of the central galaxy, the central galaxy will cover different
fractions of area that are close to the super-critical region 
(region with super-critical surface density) of the cluster.
This will modulate the cross section depending on the coverage of the cluster
super-critical region.

In addition, we have observed that the more massive is the cluster to
which the galaxy is added, the more dramatic the effect.
Throughout our experiments we have found that the ability of a cluster
to form giant arcs depends upon the number of pixels above critical in
a contiguous region near the center.
Adding a central galaxy to a high mass cluster brings the density from 
marginally critical to above critical over a wide region, thus increasing
the formation of arcs for the more massive clusters.
One might imagine that adding a galaxy to a low mass cluster would have
a large effect, bringing a region from sub- to super-critical.
However, since the cluster is low in mass it is difficult to bring a
large contiguous region above the critical density, inhibiting efficient
arc formation.

\section{Giant arcs} \label{sec:multiarc}

\subsection{The appearance of arcs}

%CHANGED
The arc distribution averaged over all the $32$ clusters, each with $32$
projections, appears to be primarily elliptical centered on the center
of the clusters.  The elliptical shapes resemble the critical curves of the
elliptical clusters, and the centers of the arcs trace out the critical curves
of the mass distribution.  The width of the arcs depends on the source
size, with sources smaller than $1''$ giving arcs which appear more similar
to those seen in optical images.

\begin{figure}
\begin{center}
\resizebox{3.5in}{!}{\includegraphics{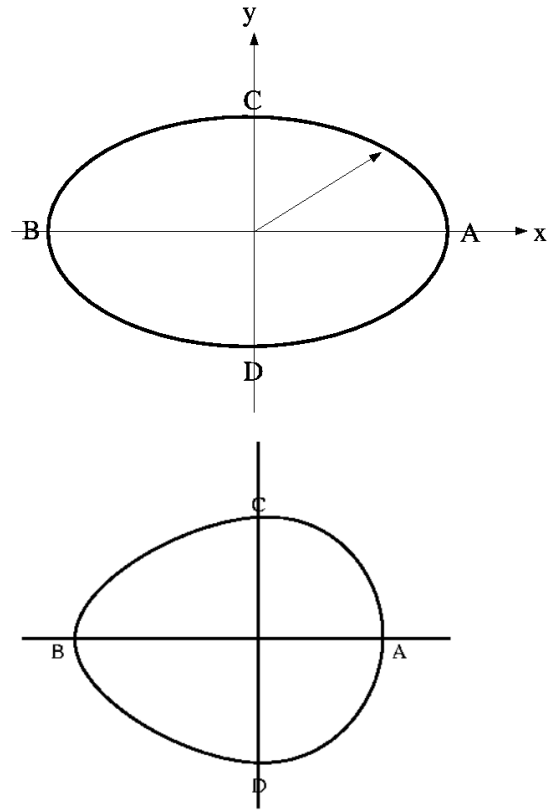}}
\end{center}
\caption{The ellipses used for describing the critical region of the lens.
The figure above describes the critical region as a ellipse, the figure below
describes the critical region as a distorted egg-shaped ellipse.} 
\label{fig:ends}
\end{figure}

As one realizes from simulations that most clusters are elliptical, and that their
critical regions (regions enclosed by their critical curves)
can be nicely fit by ellipses. We will discuss the appearance of arcs 
using terminologies that apply to ellipses to describe the lens.
Please refer to upper figure in Fig.\ref{fig:ends} for the following discussion.
Our arcs tend to form around points A, B , C and D of the elliptical 
critical region of the lens 
as this is also seen by Dalal et al.~\cite{Dalal}.

The clusters with a more circular
critical region tend to have more isotropically distributed arcs, while
those with narrower critical regions give rise to arcs that reside mainly 
around points A and B of the critical regions.

Since clusters are mostly not perfect ellipse, there are clusters with
distorted elliptical critical regions.
When there is a curvature difference between point A and B (such as the 
lower figure in Fig.\ref{fig:ends}), 
arcs are more likely to be found around point A (B) 
if the curvature around point A (B)  is smaller than around point B 
(A). 

%ADDED

This can be understood using a simple argument that there is a larger 
strip of critical curves that is available for arcs to reside and still 
be distinguished as unique arcs on the side
with less curvature than the side with larger curvature.

\subsection{Multiple arc systems}

Multiple arcs are highly beneficial to determining the structure of
clusters, and there appears to be more multiple arc systems than one
would naively expect (e.g.~Gladders et al.~\cite{Gladders}): for example,
the RCS survey recently found that 2 out of their 5 arc forming clusters
showed multiple arcs.
Therefore, it is interesting to understand the prevalence of multiple 
arcs system and to add to our understanding of how the structure of clusters
determine the formation of one or more arcs.

Using our pure dark matter maps and considering only 2 lens redshifts and 1
source redshift, we found 17 multiple arcs system
out of 31 systems (with sources at $z=1$ and the lens at $z=0.3$ and $z=0.5$)
that show arcs of L/W $>=7.5$. If we consider only the lensing systems that 
show the giant arcs (L/W $>=10$), we have 
4 multiple giant arcs systems out of 9 giant arcs systems. 
 
With a central galaxy, of mass $10^{13}\,h^{-1}M_{\odot}$, added to the
center of the projected density maps and with sources at $z=1.0$ 
, we increase the number of arcs, but the fraction of multiple arc
systems remains roughly unchanged (40-50\%). When the lensing 
efficiency of the clusters increases (with increasing source redshifts),
the fraction of multiple arc systems systematically 
increases. (see Table \ref{tab:lensingtab} for 
more statistics). 

\begin{table}
\begin{center}
\begin{tabular}{c|c|c|c}
Maps Description  & $L/W_{min}$   &$z_{src}$     & Multiple Arc    \\
                  &               &              & Fraction        \\
\hline
DM                & 10             & 1           &        4/9       \\         
DM                & 7.5            & 1           &       17/31      \\    
DM+cD             & 10             & 1           &       12/30      \\  
DM+cD             & 7.5            & 1           &       38/96      \\ 
DM+cD             & 7.5            & 2           &       49/70      \\
DM+cD             & 7.5            & 3           &       109/140    \\
DM+cD             & 7.5            & 4           &       121/156    \\
\end{tabular}
\end{center}
\caption{The multiple arc fraction (the ratio of number of 
multiple lensing systems
to number of lensing systems) for different density maps. 
The first two row describes the result from dark matter only maps at $z=0.3$
and $z=0.5$.
The remaining rows uses dark matter maps with an addition 
of a cD galaxy with maps at $z=0.3$. We also realize that with increasing 
lensing efficiency (at higher source redshifts), the ratio of number 
of multiple lensing systems increases.}
\label{tab:lensingtab}
\end{table}

Conversely, cutting out all
arcs within $10''$ of the cluster center decreases the number of arcs
but does not decrease the fraction of the multiple arc forming systems.

Finally, for a subset of the most massive clusters we verified that source
size didn't change the multiple arc fraction significantly.  For sources
from $0.2''$ to $1''$ the fraction decreased by only 8\%.
Also, the dependence of our multiple arc criterion on arc width is known
to be weak.
We infer from this that the fraction of multiple arc systems is relatively
insensitive to the details of our modeling and can be calculated reasonably
robustly using simulations such as ours.

Within the limited statistics, the fraction of multiple arc systems in our
simulations agrees with the RCS observations\footnote{We have not tried to
match the lensing rate seen by RCS, because there are numerous factors
(e.g.~source redshift dependence, size distribution, addition of central
galaxies) that can affect the rate.}.
The arc geometries are also not too dissimilar, although the statistics in
both our simulations and the observations are too poor to allow any strong
statements to be made in this regard.
The arc thickness depends on the source size and we would likely need to
reduce our sources below $1''$ to get good visual agreement, but other
properties are less sensitive to our modeling.
We infer that the high percentage of multiple arc systems seen in RCS is not
unexpected in a $\Lambda$CDM cosmology.  The good agreement may be fortuitous,
or it may be because both RCS and our simulations focused on massive clusters.

To understand the prevalence of multiple arcs we compute the arc
multiplicity function.  This lists the number of maps which have $N$
unique arcs.  To compute this we threw many sources for each map and
kept track of how many unique arcs were produced from those sources.
Unfortunately we could find no unambiguous definition of ``unique arcs''.
Slight shifts in source position lead to arcs whose properties change
continuously, making any clean separation difficult.  
To make progress we plotted the separations of different arcs and noted a
drop in the number of arcs with distance at $3.4''$, at $z=0.3$.  We picked
this distance as a cut on whether an arc produced was just a repetition of
the arcs that were produced before, or was a new arc.
Our criterion for ``unique arcs'' was that the centers be separated by
more than $3.4''$ at $z=0.3$.  With this definition, the statistics of
unique arcs seemed relatively insensitive to arc properties, such as width,
and unique arcs usually came from distinct sources in the source plane,
as seen in observations.

\begin{figure}
\begin{center}
\includegraphics[width=3.5in]{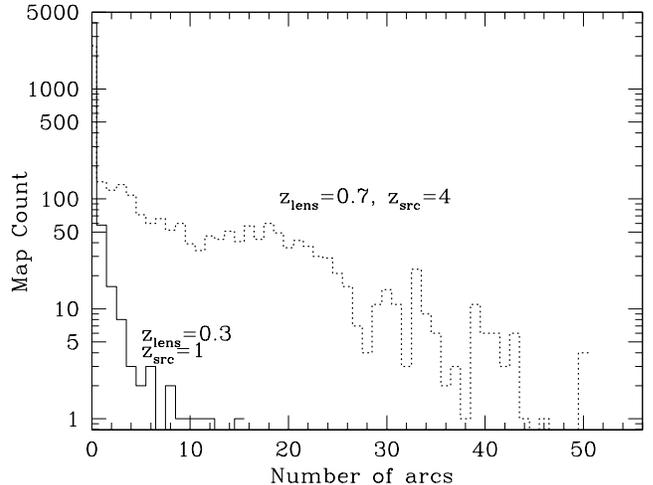}
\end{center}
\caption{The arc multiplicity function: the number of maps which showed
%CHANGED
$N$ unique arcs  of L/W $>=7.5$  (see text) when a large number of sources were thrown.
The dotted line indicates the multiplicity function for clusters at
$z=0.7$ with sources at $z=4.0$, the solid line shows the multiplicity
function for clusters at $z=0.3$ with sources at $z=1.0$.}
\label{fig:multi}
\end{figure}

The multiplicity function is shown in Figure \ref{fig:multi}.
Note that the vast majority of maps show no giant arcs.
Among systems which show at least one arc however, a significant subset show
multiple arcs.  Note the extreme tails to this distribution, with maps that
could in principle host tens of giant arcs were the sources properly aligned.
This suggests that clusters in a $\Lambda$CDM cosmology could provide a
rich array of arc possibilities, even before we consider substructure in
the galaxies being lensed.
While the probability that a cluster will host an arc, $p$, could be very
small, once a lens is massive enough to host one arc the probability that
it hosts a second is not suppressed by another power of $p$
(c.f.~Gladders et al.~\cite{Gladders}).

\begin{figure}
\begin{center}
\includegraphics[width=3.5in]{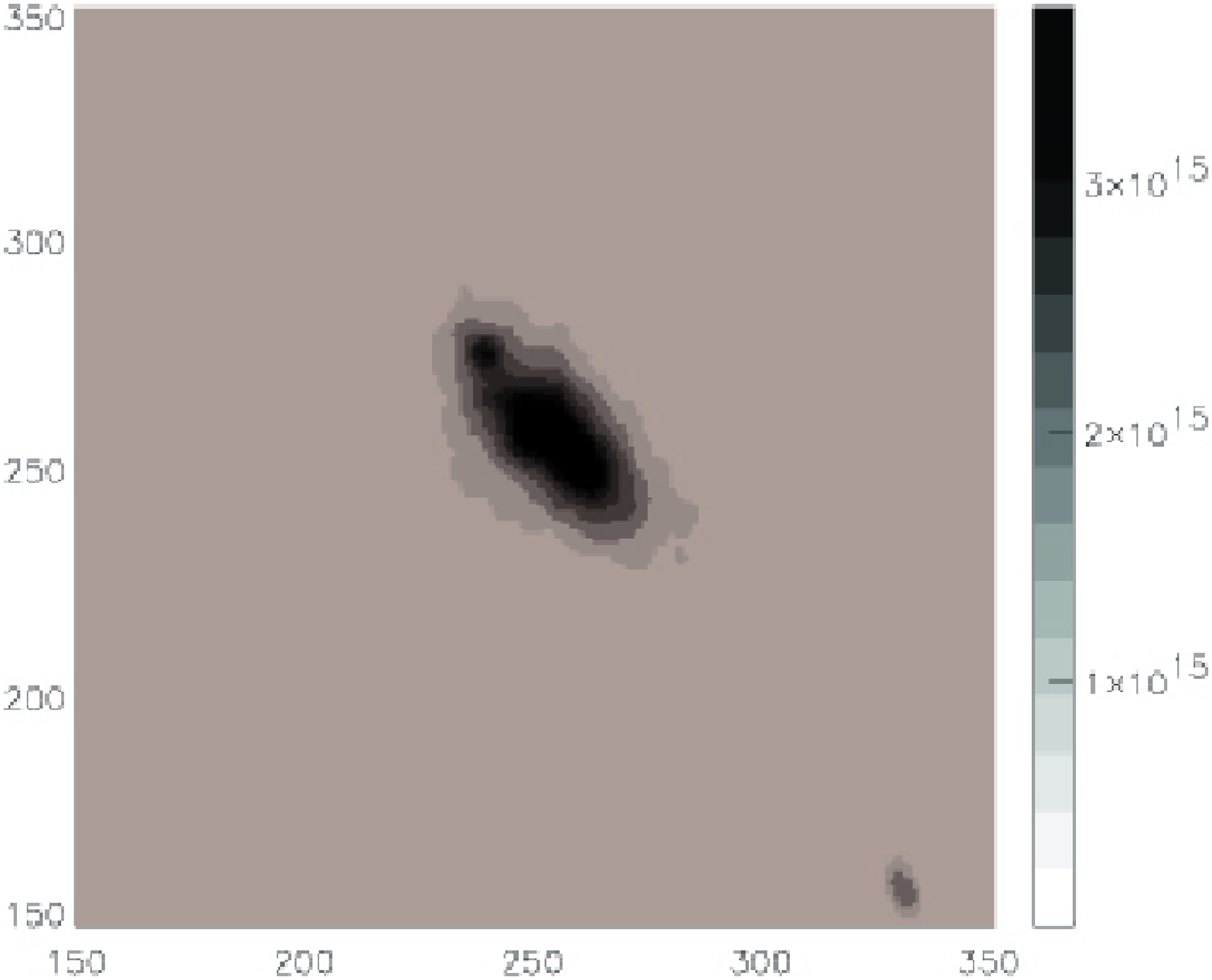}
\includegraphics[width=3.5in]{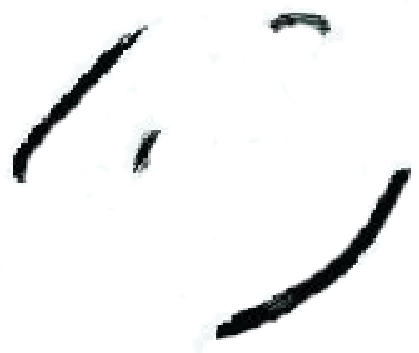}
\end{center}
\caption{(Upper) The projected mass (in $h^{-1}M_\odot/(h^{-1}{\rm Mpc})^2$)
of the super-lensing cluster viewed in the plane where the nearby structure
resides.  The $x$- and $y$-axes are in pixels of size $9.77\,h^{-1}$kpc.
(Lower) One example of a multiple arc system formed by this cluster.}
\label{fig:clus}
\end{figure}

Among the $32$ clusters at $z=0.3$, we also find one ``super-lensing''
cluster.  Three of the projections of this cluster, of the $4096$ total maps,
produced $\sim 90\%$ of the arcs produced when we placed the sources at
$z=1.0$.
To understand this behavior we looked at the properties of the cluster in
some detail.  It is very massive,
$M_{200}\simeq 1.5\times10^{15}\,h^{-1}M_{\odot}$,
and has its nearby structure residing almost in a plane
(see Figure \ref{fig:clus}).
When the line of sight is parallel to this plane not only the cluster but
its neighbors and connecting filaments contribute to the projected mass
``at'' the lens, leading to a very efficient lensing configuration.
The other orientations are not as efficient.

It is interesting that the arc pattern of this super-lensing cluster is
similar to the system RBS653 analyzed by Kausch et al.~\cite{Kausch},
with multiple giant arcs formed at the two ends of the 
cluster. 
We do not have a large sample of such very massive, efficient clusters due
to the limited volume we have simulated.
It is not unreasonable to expect that observations, probing larger volumes,
are picking up massive clusters which are highly efficient lenses, being
surrounded by a large amount of correlated structure.

\subsection{Mass distributions that are efficient in lensing}

The existence of the super-lensing cluster leads us to ask: when would a mass
distribution be a good lens?  To answer this question we have taken our sample
of simulated clusters and looked at the projected mass density of $\sim 20$
of them.  They are all of similar mass and some are efficient lenses while
the rest are not.

We noticed that the efficient clusters had reasonably large contiguous
regions above $\Sigma_{\rm crit}$, while the less efficient clusters
had separated regions above $\Sigma_{\rm crit}$. 
A reasonably sized contiguous region above $\Sigma_{\rm crit}$ typically
has more sources which can be lensed sufficiently to form a giant arc.
Looking at this from another viewpoint, we can also say that a larger 
contiguous super-critical region would mean the caustic extends over 
a larger region, thus giving rise to more lensing events in which
sources (probably farther from 
each other) are
forming arcs that are
more likely to be apart in the image plane, thus easier to be uniquely 
identified. On the other hand, two separated regions (imagine we cut 
the previous contiguous region into two) 
above $\Sigma_{\rm crit}$
will then give rise to lensing events that consist of several 
sources very close to each other being lensed and form images that 
are fairly close to each other, 
thus harder to be distinguished as separate arcs,
lowering the number of the arcs formed.
This illustrates the fact that arcs merge complicates 
the correlation between the length of the caustic and the number 
of arcs that can be produced.

This observation also explains why the arc statistics are so sensitive to
source redshift but less sensitive to central galaxies (apart from 
contribution from the
change of source size due to changing source redshift).
Imagine the cluster has a rather smooth mass distribution.  Increasing
$z_{\rm src}$ decreases $\Sigma_{\rm crit}$ and enlarges the region
above critical density, giving rise to many more arcs.
By contrast adding a central galaxy only brings the region near the center
above critical, which may not be contiguous with other super-critical
regions in the map.

\section{Discussion and Conclusion} \label{sec:conclusions}

Recent observations have begun to amass statistics on giant arcs around
clusters of galaxies.  Such arcs probe the mass distribution of clusters
on scales which should be amenable to theoretical interpretation (see
Appendices), making lensing arcs a new meeting ground between theory and
observation.
This motivated us to consider how giant arcs are formed, and how one should
use arcs to probe underlying clusters.
To further our understanding in how arcs are formed, and how
arcs can be used as a mass probe, we used ray tracing through N-body
simulations to make simulated images of giant arcs.
In addition to the dark matter followed by the simulations we have added
analytic profiles representing central galaxies of various masses,
orientations and ellipticities, used different source redshifts, source 
sizes and clusters
at different redshifts to investigate the giant arcs statistics.  

Our work is not the first to try understanding how arcs are formed. 
Where there is overlap our work agrees with some of the earlier simulations.
In particular we agree with Dalal et al.~\cite{Dalal} on the effects of
central galaxies and orientation of the cluster on the arc formation.
We also agree with Wambsganss et al.~\cite{Wambs} on the strong dependence
of arc cross section on source redshift, though we do not have enough
information from their paper to make a precise comparison.
We further the investigation of the strong redshift dependence, 
realized that with increasing source redshift, the cross section increases
more drastically with constant physical source size than with constant 
angular source size.  
This redshift dependence may explain why RCS finds giant arcs in predominantly
high-$z$ clusters.
We agree with recent work suggesting that the link between the number of
giant arcs and cosmology is complex, with assumptions about central
galaxies, source redshifts, sizes, ellipticities and substructures making
large differences in the theoretical predictions.  We find in particular,
and contrary to the claim by Bartelmann et al.~\cite{Bart1}, that the
lensing cross section does depend on the source size assumed.
This comes about through the selection of arcs above some fixed $L/W$ ratio,
since the smaller the source size the narrower arc that is produced.

We investigated why some clusters are more effective lens than the others.
We found a correlation between the area of the contiguous region above
$\Sigma_{\rm crit}$ and the arc cross section.
This helps to explain why certain clusters are better arc producers than
other, comparable mass, clusters and why adding central galaxies is not
as efficient as increasing the source redshift.      

We have found one cluster in our simulations that is an extremely efficient
lens.  Given our relatively small simulation volume this suggests that
CDM cosmologies should naturally produce clusters which are efficient
arc, or even multiple arc, producers. 
Given the prevalence of multiple arcs systems in various observations,
we investigated the multiple arcs system by looking at the number
of unique arcs that arise from ray tracing our cluster sample.
The fraction of systems producing
multiple arcs in our simulations was quite insensitive to the details of
our modeling, and close to the fraction seen in observations
(Gladders et al.~\cite{Gladders}).

Looking forward into the future, upcoming wide-field observations such as
RCS{\sc ii}\footnote{www.astro.utoronto.ca/~gladders/RCS},
the CFHT Legacy Survey\footnote{http://www.cfht.hawii.edu/Science/CFHLS},
the Sloan Digital Sky Survey\footnote{http://www.sdss.org} 
and X-ray cluster surveys such as
MACS\footnote{http://www.ifa.hawaii.edu/~ebeling/press/macs/images.html}
can be expected to improve the statistics of giant arcs on the sky.
With this increase in statistics we expect giant arcs will be an
indispensable probe of the structure of clusters and the formation of
large-scale structure.

\begin{acknowledgments}
S.H would like to thank our anonymous referee for giving us very
interesting suggestions and corrections. S.H would also like to thank 
Henk Hoekstra, Neal Dalal, Niayesh Afshordi
 and the Berkeley Cosmology Group for
discussions and Joanne Cohn for both her help and insightful discussions.
The simulations used here were performed on the IBM-SP at the National
Energy Research Scientific Computing Center.
This work was supported by grants from the NSF and NASA.
\end{acknowledgments}

\appendix

\section{Comparison with analytic profiles}

There is a large literature surrounding the theory of strong gravitational
lensing, much of it based on simple analytic potentials
(e.g.~Narayan \& Bartelmann~\cite{Nara}, Keeton~\cite{Keeton},
Kochanek et al.~\cite{Kochanek} and references therein).
In order to understand how well such potentials describe our simulated
clusters we compare here the critical curves and caustics of a select few
of our clusters with those of the analytic form
\begin{equation}
  \Sigma = {\Sigma_0\over r(1+r^2)}
\end{equation}
%CHANGED
where $r=\left(qx^2+[y]^2/q \right)^{1/2}/R_c$.
This analytic surface density is of a broken power-law form with finite mass,
here taken to be $10^{15}\,h^{-1}M_{\odot}$.  We set the scale radius to
$R_c=70\,h^{-1}$kpc and the ellipticity ($q$) to $0.7$.

Arcs should be formed when the sources cross the caustics on the source plane
where formally the magnification becomes infinite.  The arcs trace out the
critical curves in the image plane, being highly elongated and magnified
images of the critical source.

We show the magnification of this analytic potential as a function of
%CHANGED
position in the source plane in Fig 12.  Note the two `lips'
%position in the source plane in Fig \ref{fig:analyt}.  Note the `diamond'
shaped structure commonly seen in potentials of this form.
We can compare this to the same plot for two clusters from our N-body
simulation (Fig 13).
%simulation (Fig \ref{fig:real19}).
The first cluster is the ``super-lensing'' cluster (Fig.~\ref{fig:clus})
of \S\ref{sec:multiarc}.  Note the ``squeezed diamond'' shape of the
magnification distribution for this cluster, which we are seeing projected
edge-on.  A less extreme example is given in the other panel of
Fig 13.
%Fig \ref{fig:real19}.

We have found that the longer the caustic, the larger the cross section
for lensing.  This can be understood from the fact that there is a higher
probability of the sources being largely magnified when we have a longer
caustic.  This is dramatically illustrated here: the cluster with shorter
and smaller caustic (second panel) does not lens nearly as well as the
cluster with the longer and larger caustic (first panel).

\begin{figure}
\begin{center}
\includegraphics[width=3.5in]{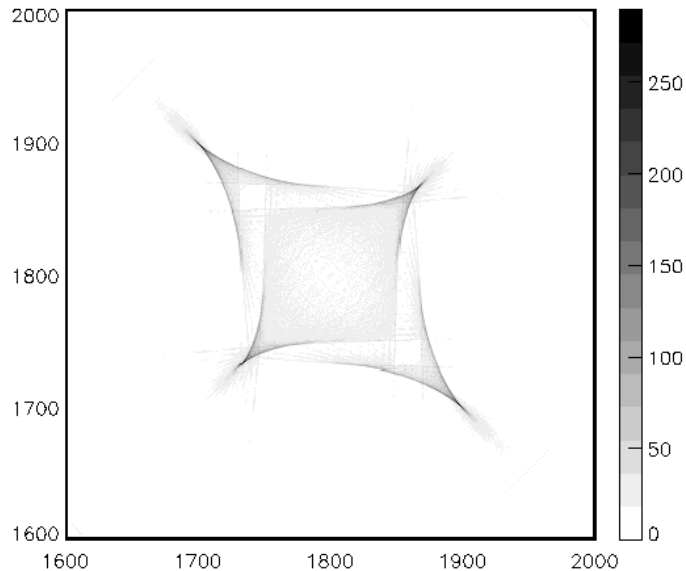}
\label{fig:analyt}
\end{center}
\caption{A greyscale image of the magnification of the analytic potential
as a function of position on the source plane (in pixels).  Note the 
%CHANGED
two lips shaped caustic structure.}
\end{figure}

\begin{figure}
\begin{center}
\includegraphics[width=7.0in]{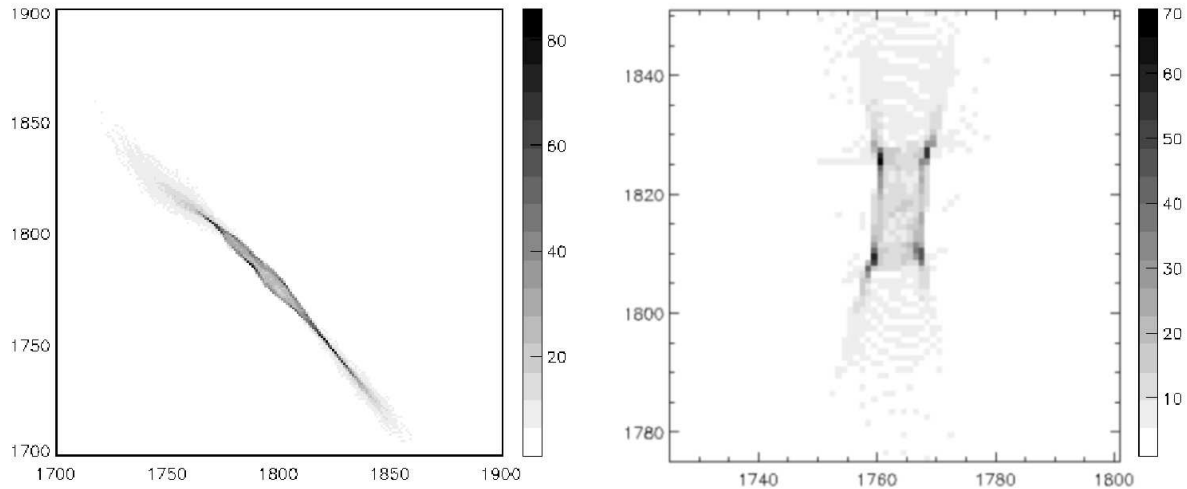}
\label{fig:real19}
\end{center}
\caption{Greyscale images of the magnification, as a function of position in
the source plane (in pixels), of the super-lensing cluster described
previously (left panel) and a cluster not being viewed edge-on (right panel).
Again the caustic structure is clearly visible.}
\end{figure}

\section{Einstein Radius}

The relevant scale for gravitational lensing is the Einstein radius, the
radius of the circle formed by a spherically symmetric mass distribution
lensing a perfectly on-axis source.  It is useful to work through the
calculation of the Einstein radius for the case of interest here.
Solving the lens equation for an on-axis source being lensed by a
spherically symmetric mass distribution we find
\begin{equation}
  r_E = 4 \Delta  \frac{G M(<r_E)}{c^2 r_E}
\label{eqn:einstein}
\end{equation}
where $M(<r)$ is the mass enclosed within radius $r$, $\Delta=D_{LS}D_L/D_S$
and $D_L$, $D_{LS}$ and $D_S$ are the angular diameter distances between the
observer and the lens, the lens and the source, and the observer and the
source respectively.
If we define the (proper) gravitational radius of the lens as
\begin{equation}
  r_g = {4G M_{\rm tot}\over c^2}
\end{equation}
we can rewrite Eq.~\ref{eqn:einstein} as
\begin{equation}
  \left({r_E\over R_*}\right)^2  = {M(<r_E)\over M_{\rm tot}}
\label{eqn:re}
\end{equation}
where $R_*\equiv\sqrt{r_g\Delta}$.  For a $10^{15}\,h^{-1}M_\odot$ lens
at $z\simeq 0.3$ and sources at $z\simeq 1$ the characteristic scale
$R_*\simeq 300\,h^{-1}$kpc comoving.  The scale is relatively insensitive
to the source and lens redshift, within the interesting range $R_*$ lies
between $300$ and $450\,h^{-1}$kpc.  Typically $r_E<R_*$.

To solve Eq.~\ref{eqn:re} we need to assume a profile $M(<r)$.
Unfortunately for an NFW profile the solution is quite unstable, because
$M(<r)\sim r^2$ for $r\to 0$.  Small changes in the assumed mass or
lensing geometry can dramatically alter the solution for $r_E$.
This is one of the reason why the ability of clusters to form arc
has such a strong dependence on e.g. the viewing orientation.
Physically this extreme sensitivity is mitigated by the presence of a
central cusp in the mass distribution which is steeper than $\rho\sim r^{-1}$
for $r\to 0$, for example a central galaxy with an isothermal profile.
Thus for many systems the defining scale is set by the radius at which
baryonic cooling has steepened the central profile beyond $1/r$.  The
Einstein radius is then determined by the galactic radius and is of
order $100$kpc.

\end{document}